\documentclass[pre,twocolumn,groupedaddress,showpacs]{revtex4}
\usepackage{graphicx}
\usepackage{dcolumn}
\usepackage{amsmath}
\usepackage{bm}
\begin{document}
%\draft
\title{Liquid crystal director fluctuations and surface anchoring by
molecular simulation}
\author{Denis Andrienko$^1$} 
\author{Guido Germano$^{1,2}$}
\author{Michael P. Allen$^1$}
\affiliation{$^1$ H. H. Wills Physics Laboratory, University of Bristol \\
Royal Fort, Tyndall Avenue, Bristol BS8 1TL, United Kingdom \\
$^2$ Fakult\"{a}t f\"{u}r Physik, Universit\"{a}t Bielefeld,
33501 Bielefeld, Germany}
\begin{abstract}
  We propose a simple and reliable method to measure the liquid crystal surface
  anchoring strength by molecular simulation. The method is based on the
  measurement of the long-range fluctuation modes of the director in confined
  geometry. As an example, molecular simulations of a liquid crystal in
  slab geometry between parallel walls with homeotropic anchoring have been
  carried out using the Monte Carlo technique.  By studying different slab
  thicknesses, we are able to calculate separately the position of the elastic
  boundary condition, and the extrapolation length.
\end{abstract}
\maketitle
\section{Introduction}
\label{sec:intro} 
Liquid crystal anchoring effects have been intensively
studied both experimentally and theoretically during the past decades 
\cite{jerome.b:1991.a,poniewierski.a:1996.a}. 
Such an interest is well understood since
most liquid crystal devices are cells comprising orienting surfaces with a
liquid crystal in between. 
Typically, aligning surfaces provide a uniform orientation
of the liquid crystal director in the cell interior.

On the phenomenological level, 
liquid crystal anchoring can be described by two basic parameters: 
the easy axis direction, ${\bf e}$, 
and the anchoring coefficient $W$. 
These two parameters are critical design parameters for
every liquid crystal device 
\cite{andrienko.d:2000.a}.
A variety of {\em experimental} methods have been proposed to measure
anchoring parameters, 
in particular the anchoring coefficient $W$ 
\cite{gu.df:1995.a,andrienko.d:1998.a}. 
Most of them measure surface director deviations in an external field 
and involve rather complicated optical setups.

In spite of the practical importance, 
the mechanism of the director alignment is still not well understood. 
Experimental techniques always involve optical measurements. 
They test the entire liquid crystal cell and
therefore cannot provide a satisfactory description 
of the thin interface region. 
Theoretical investigations of anchoring are also quite controversial.
For example, 
the usual continuous phenomenological theory has divergent surface terms
in the elastic free energy expansion 
\cite{pergamenshchik.vm:1993.a,pergamenshchik.vm:1999.a}.

One of the approaches 
for the systematic investigation of anchoring phenomena
is computer simulation of liquid crystals in confined geometries.
Indeed, computer simulation is a well established method 
to treat bulk elastic coefficients 
\cite{1988.a,1990.a}, 
the surface anchoring strength 
\cite{1999.b}, 
and structures of disclination cores 
\cite{hudson.sd:1993.a,2000.c}. 
This means that computer simulation allows investigation 
of details of the liquid crystalline structure 
which cannot be resolved experimentally.

Several papers have already been published 
trying to search for reasonable surface potentials 
to use in simulations 
\cite{zhang.z:1996.a,stelzer.j:1997.c,stelzer.j:1997.b,%
gruhn.t:1997.a,gruhn.t:1998.a,wall.gd:1997.a}. 
However, most of them do not characterize aligning surfaces 
using well established parameters. 
Questions about the formation of a solid interface layer, 
values of the easy axis angle and anchoring coefficient 
are still open. 
The reason for this is probably a lack of reliable methods to measure
these parameters by computer simulation.

In this paper, 
we propose a technique to measure the surface anchoring strength $W$ 
by computer simulation. 
The method itself is based on 
the study of the {\em director fluctuations} 
in the liquid crystal cell. 
A similar approach has already been used 
for the {\em experimental} characterization of the interface 
\cite{marusii.ty:1986.a} 
and is based on measuring the light scattering 
caused by the director fluctuations.
In computer simulation, 
the director fluctuations can be studied directly,
using ensemble averages of functions of 
the second-rank order tensor components. 
A fit to the fluctuation amplitudes 
with equations predicted by elastic theory 
then allows determination of the surface anchoring strength.
\section{Theory}
\label{sec:theory}
Large length- and time-scale fluctuations of the director ${\bf n}({\bf r})$ 
can be described in the continuum model of liquid crystals,
based on the phenomenological elastic constants. 
In this approach, 
the hydrodynamic equations for the director and the boundary conditions 
can be obtained by minimization of the cell free energy 
\cite{degennes.pg:1995.a}: 
\begin{equation}
F=F_{\rm h}+F_{\rm b}+F_{\rm s} \:,
\label{eqn:f_total}
\end{equation}
where 
$F_{\rm h}={{\frac{1}{2}}}\gamma (\partial{\bf n}/\partial t)^{2}$ 
is a hydrodynamic term with an effective viscosity coefficient $\gamma $, 
$F_{\rm b}$ is the Frank elastic free energy, 
and $F_{\rm s}$ is the surface anchoring energy.

In what follows we use the one-elastic-constant approximation, 
i.e.\ $K_{11}=K_{22}=K_{33}=K$. 
Then the Frank free energy can be brought into the form: 
\[
F_{\rm b}
=
\frac{1}{2}K\int_V \left[ 
\left( \nabla \cdot {\bf n}\right) ^{2}
+
\left(\nabla \times {\bf n}\right) ^{2}
\right] d{\bf r} \:.
\]

The liquid crystal cell is bounded by surfaces $z=0,L$ 
which provide some kind of anchoring condition 
\cite{jerome.b:1991.a}. 
Below we consider {\em homeotropic} anchoring, 
that is, normal to the surface. 
{\em Planar} anchoring can be treated in the same way.
We assume that the interaction of the director with the cell surfaces 
has the Rapini-Papoular form 
\cite{rapini.a:1969.a}: 
\[
F_{\rm s}
=
-\frac{1}{2}W\int_{S_{0},S_{L}}~
\left( {\bf n} \cdot {\bf e}_{0,L}\right)^{2} d{\bf r}_{\perp },
\]
where unit vectors ${\bf e}_{0,L}$ 
define directions of the easy axes at 
$z=0,L$: ${\bf e}_{0}={\bf e}_{z}$, ${\bf e}_{L}=-{\bf e}_{z}$; 
$W$ is the anchoring energy, 
and ${\bf r}_{\perp }=(x,y)$.

Equations for the director and boundary conditions 
can be obtained by minimization of the cell free energy 
eqn~(\ref{eqn:f_total}). 
In the case of homeotropic anchoring on both surfaces 
the stationary director distribution in the cell is homeotropic, 
i.e.\ ${\bf n}_{0}={\bf e}_{z}$. 
Therefore, 
we have to investigate small perturbations of the director 
around the distribution: 
\begin{equation}
{\bf n}({\bf r}) ={\bf n}_{0}+\delta {\bf n}( {\bf r}) \:.  
\label{eqn:director}
\end{equation}
Minimizing the total free energy (\ref{eqn:f_total}) 
and linearizing the equations for the director and boundary conditions 
with respect to $\delta {\bf n}$, 
we obtain equations 
\begin{equation}
\gamma \frac{\partial }{\partial t}\delta {\bf n}
=
K\nabla ^{2}\delta {\bf n} \:,
\label{eqn:hydro}
\end{equation}
and boundary conditions
\begin{equation}
\left. 
W\delta {\bf n\pm }K\frac{\partial }{\partial z}\delta {\bf n}
\right|_{z=L,0}
=
{\bf 0} \:.
\label{eqn:boundary}
\end{equation}
for the fluctuations.
Taking into account eqn.~(\ref{eqn:director}),
the expression for the free energy fluctuations 
can be rewritten in the form of the average of the
self-conjugate (Hermitian) operator $-\frac{1}{2}K\nabla ^{2}$: 
\begin{equation}
\delta F_{\rm b}
=
-\frac{1}{2}K\int_V ~
\delta {\bf n}({\bf r}) 
\nabla^{2}\delta {\bf n}({\bf r}) ~d{\bf r} \:.  
\label{eqn:denergy}
\end{equation}
The eigenfunctions of the operator $-\frac{1}{2}K\nabla ^{2}$, 
which satisfy boundary conditions (\ref{eqn:boundary}), 
form a complete set of orthogonal functions 
characterized by wave vector ${\bf q}$. 
Therefore, $\delta {\bf n}({\bf r})$ 
can be expanded in a series of these orthogonal functions: 
\begin{eqnarray}
\nonumber
\delta {\bf n}( {\bf r}) 
&=&
\frac{1}{V}\sum_{{\bf q}_{\bot},q_{z}}
{\rm e}^{{\rm i}{\bf q}_{\bot}\cdot{\bf r}_{\bot}} \times
\\
&\times& \left[ 
\delta {\bf n}^{(+)}({\bf q}_{\bot},q_{z}){\rm e}^{{\rm i}q_{z}r_{z}}
+
\delta {\bf n}^{(-)}({\bf q}_{\bot},q_{z}){\rm e}^{-{\rm i}q_{z}r_{z}}
\right] \:,  
\label{eqn:fluct}
\end{eqnarray}
where ${\bf q}_{\bot}=(q_{x},q_{y})$, and 
\begin{equation}
\delta {\bf n}^{(-)}=\frac{{\rm i}\chi-\xi}{{\rm i}\chi+\xi}\delta {\bf n}^{(+)} \:.  
\label{eqn:nplusminus}
\end{equation}
Here we have introduced the dimensionless wave vector $\chi =q_{z}L$ 
and anchoring parameter 
\begin{equation}
\xi =\frac{WL}{K}=\frac{L}{\lambda } \:,  
\label{eqn:anchoring}
\end{equation}
where $\lambda$ is the extrapolation length
\cite{degennes.pg:1995.a}.
The wave vectors $q_{z}$ form a discrete spectrum 
because of confinement in the $z$ direction 
which depends on the anchoring energy $W$ 
(see Appendix for details). 
The explicit form of the $q_{z}$ spectrum is given by the secular equation: 
\begin{equation}
\left(\xi^{2}-\chi^{2}\right) \sin\chi +2\xi\chi \cos\chi =0 \:.
\label{eqn:secular}
\end{equation}
Each individual mode can now be seen from eqn~(\ref{eqn:hydro}) 
to relax exponentially with a relaxation time given by 
\begin{equation}
\tau =\gamma /K\left( {\bf q}_{\bot }^{2}+q_{z}^{2}\right) \:.  
\label{eqn:time}
\end{equation}

Substituting expansion (\ref{eqn:fluct}) 
into the free energy (\ref{eqn:denergy}),
and performing the integration over the cell volume we obtain: 
\begin{eqnarray}
\nonumber
\delta F_{\rm b}
&=&
\frac{K}{V}\sum_{{\bf q}_{\bot},q_{z}}
\frac{q^{2}\left(2\xi+\chi^{2}+\xi^{2}\right)}{({\rm i}\chi+\xi)^{2}}
\times \\
&\times& \delta {\bf n}^{(+)}( {\bf q}_{\bot },q_{z}) 
\delta {\bf n}^{(+)}(-{\bf q}_{\bot },q_{z}) \:.
\end{eqnarray}
Integrating, 
we took into account the orthogonality of the eigenfunctions 
in the expansion (\ref{eqn:fluct}) 
with different eigenvectors ${\bf q}$, 
which allowed us to reduce the summations over ${\bf q}$, ${\bf q}^{\prime}$ 
to a single sum over ${\bf q}$.

Application of the equipartition theorem of classical statistical mechanics,
just as for elastic fluctuations in bulk liquid crystals 
\cite{forster.d:1975.a}, 
gives the fluctuation amplitudes: 
\begin{eqnarray}
\nonumber
\left\langle 
\delta {\bf n}^{(+)}( {\bf q}_{\bot},q_{z}) 
\delta {\bf n}^{(+)}(-{\bf q}_{\bot},q_{z}) 
\right\rangle 
= \\
-\frac{k_{{\rm B}}TV({\rm i}\chi+\xi)^2}{2Kq^2\left(2\xi+\chi^2+\xi^2\right)} 
\:,
\end{eqnarray}
where $\left\langle\ldots\right\rangle$ denotes an ensemble average.

In molecular simulations,
rather than measuring director fluctuations, 
it is more convenient to measure fluctuations of the 
second-rank order tensor components 
(following Forster 
\cite{forster.d:1975.a}).
We define the real-space order tensor density 
\begin{eqnarray*}
Q_{\alpha\beta}({\bf r}) &=&\frac{V}{N}\sum_{i}
\delta({\bf r}-{\bf r}_{i}) Q_{\alpha\beta }^{i} 
\:, \\
Q_{\alpha\beta}^{i} &=& 
\frac{3}{2}\left(u_{i\alpha}u_{i\beta}-\frac{1}{3}\delta_{\alpha\beta}\right)
\:,
\end{eqnarray*}
where $\alpha ,\beta =x,y,z$,
in terms of the orientation vectors 
${\bf u}_{i}$ of each molecule $i$ (we consider only uniaxial molecules).
If we assume that there is no variation in the {\em degree} of ordering,
we may write
\[
Q_{\alpha\beta}({\bf r}) = 
\frac{3}{2}Q n_{\alpha}({\bf r})n_{\beta}({\bf r}) 
- \frac{1}{2}Q\delta_{\alpha\beta}
\]
where $Q$ is the order parameter, 
i.e.\ the largest eigenvalue of $Q_{\alpha\beta}({\bf r})$.
If the director ${\bf n}_{0}$ 
is taken to lie along the $z$ axis throughout the sample,  
the off-diagonal components 
$Q_{\alpha z}({\bf r}) $, $\alpha =x,y$,
are proportional to the
fluctuations of the corresponding director components
\[
Q_{\alpha z}({\bf r}) =\frac{3}{2}Q\delta n_{\alpha }( {\bf r}) \:.
\]
This is the situation for homeotropic anchoring at both surfaces. 
For planar anchoring, 
with the director along $x$ and the surface normal along $z$, 
the components $Q_{xy}$, $Q_{xz}$ are important (and non-equivalent).

\begin{widetext}

Measurements are performed directly in reciprocal space. 
The Fourier transform of the real-space order tensor is 
\[
Q_{\alpha\beta}({\bf k}) =
\int_V ~ Q_{\alpha\beta}({\bf r}) {\rm e}^{{\rm i}{\bf k}\cdot{\bf r}}d{\bf r}
=
\frac{V}{N}\sum_i~Q_{\alpha\beta}^{i} {\rm e}^{{\rm i}{\bf k}\cdot{\bf r}_i}
\:.
\]

Then the fluctuations 
$\left\langle\left| Q_{\alpha\beta}({\bf k})\right|^{2}\right\rangle$ 
can be easily measured from simulations: 
\begin{equation}
\left| Q_{\alpha\beta}({\bf k})\right| ^{2}
=\frac{V^{2}}{N^{2}}
\left[\left(\sum_i~Q_{\alpha\beta}^i\cos({\bf k}\cdot{\bf r}_i)\right)^2
     +\left(\sum_i~Q_{\alpha\beta}^i\sin({\bf k}\cdot{\bf r}_i)\right)^2
\right] \:.  
\label{eqn:simul}
\end{equation}

We explicitly relate simulation-measured fluctuation modes with 
theoretically predicted amplitudes of the director fluctuations
for ${\bf q}_{\bot }=0$
\[
Q_{\alpha z}(k_{z}) =\frac{3}{2{\rm i}}Q\sum_{q_{z}}~
\delta {\bf n}^{(+)}({\bf 0},q_{z}) 
\left[ 
\frac{{\rm e}^{{\rm i}(\kappa+\chi)}-1}{\kappa+\chi}
+\left(\frac{{\rm i}\chi-\xi}{{\rm i}\chi+\xi}\right)
\frac{{\rm e}^{{\rm i}(\kappa-\chi)}-1}{\kappa-\chi}
\right]
\]
where $\kappa =k_{z}L$. 
Note that the $q_{z}$ take discrete (but not equally spaced) 
values as discussed earlier, 
while the $k_{z}$ values are unrestricted.

Since $\delta {\bf n}({\bf r})$ is real, 
using eqn~(\ref{eqn:fluct}) we have 
\[
\left[ \delta {\bf n}^{(+)}({\bf q}_{\bot},q_{z})\right]^{\ast}
=-\frac{\xi^{2}+\chi^{2}}{\left( \xi +{\rm i}\chi \right) ^{2}}
\delta {\bf n}^{(+)}(-{\bf q}_{\bot},q_{z}) \:.
\]
Taking this equation into account,
and the fact that fluctuations with
different wavevectors are independent, 
i.e.\ 
$\left\langle 
\delta {\bf n}^{(+)}({\bf q}_{\bot},q_{z}) 
\delta {\bf n}^{(+)}(-{\bf q}_{\bot},q_{z}^{\prime })
\right\rangle =0$ 
if $q_{z}\neq q_{z}^{\prime}$, 
the corresponding ensemble
average of the squared order parameter can be rewritten as 
\begin{equation}
\left\langle \left| Q_{\alpha z}(k_{z}) \right|^{2}\right\rangle 
=
\frac{9}{8}k_{{\rm B}}T\frac{Q^{2}V}{K}\sum_{q_{z}}~
\frac{\chi^2+\xi^2}{q_{z}^{2}\left( 2\xi +\chi ^{2}+\xi ^{2}\right) }
\left| 
\frac{{\rm e}^{{\rm i}(\kappa+\chi)}-1}{\kappa+\chi}
+
\left(\frac{{\rm i}\chi -\xi }{{\rm i}\chi +\xi }\right)
\frac{{\rm e}^{{\rm i}(\kappa-\chi)}-1}{\kappa-\chi}
\right|^{2} \:.  
\label{eqn:theory}
\end{equation}

We measure $Q$ and
$\left\langle\left| Q_{\alpha z}( k_{z}) \right|^{2}\right\rangle$ 
from simulations, eqn~(\ref{eqn:simul}), 
and then compare with the theoretical prediction, 
eqn~(\ref{eqn:theory}),
which is parametrized by $L$, $\lambda$ and $K$.
Both the permitted $q_z$ spectrum,
and the variation of
$\left\langle\left|Q_{\alpha z}(k_{z})\right|^2\right\rangle$ 
with $k_z $ are sensitive to the anchoring strength
parameter $\xi=L/\lambda$ .

Fluctuation amplitudes given by eq. (\ref{eqn:theory}) have features
that simplify the fitting procedure. First,
terms with small $q_{z}$ values dominate because of the $q_{z}^{2}$ in the
denominator of eq (\ref{eqn:theory}). Therefore, it is always possible to
truncate this sum and use only the first values of the $q_{z}$ spectrum.
Then, for large $k_{z}$ or for $\kappa >>\chi $, eq. (\ref{eqn:theory}) can
be simplified so that the dependence on $k_{z}$ is explicit: 
\begin{equation}
\left\langle \left| Q_{\alpha z}(k_{z})\right| ^{2}\right\rangle =\frac{9}{2}%
k_{{\rm B}}T\frac{Q^{2}V}{K}\left[ \frac{\sin \left( \kappa /2\right) }{%
\kappa /2}\right] ^{2}\sum_{q_{z}}~\frac{\chi ^{2}}{q_{z}^{2}\left( 2\xi
+\chi ^{2}+\xi ^{2}\right) }.  \label{eqn:theory:approx}
\end{equation}

Therefore, for large $k_{z}$, the fluctuation amplitude $\left\langle \left|
Q_{\alpha z}(k_{z})\right| ^{2}\right\rangle $ has a characteristic
oscillation with the period given by $\kappa =k_{z}L=2\pi $. This means that
we can adjust the cell thickness $L$ independently of parameters $\lambda $ and $K$ by examining characteristic wavelength of the fluctuation amplitude $
\left\langle \left| Q_{\alpha z}(k_{z})\right| ^{2}\right\rangle $.

\end{widetext}

\section{Molecular model and simulation methods}
\label{sec:simulation}
To test the technique proposed, 
we simulated a liquid crystal confined between parallel walls (slab geometry), 
with finite homeotropic anchoring at the walls. 
The director fluctuations occur around
the preferred, {\em uniform}, alignment
perpendicular to the walls.

We performed Monte Carlo (MC) simulation of the liquid crystal system. 
We used a molecular model which has been
studied earlier in this geometry 
\cite{1999.b}. 
The molecules in this study 
were modelled as hard ellipsoids of revolution of elongation $e=a/b=15$, 
where $a$ is the length of the semi-major axis 
and $b$ the length of the two equal semi-minor axes. 
The phase diagram and properties of this family
of models are well studied 
\cite{frenkel.d:1984.a,frenkel.d:1985.a,1992.a,1993.g,1993.j}. 
%Units of length are
%chosen such that $8ab^{2}=1$, making the molecular volume equal to that of a
%sphere of unit diameter. 
It is useful to express the density as a fraction
of the close-packed density $\rho _{cp}$
of perfectly aligned hard ellipsoids, assuming
an affinely-transformed face-centred cubic or hexagonal close packed lattice. 
%in reduced units $\rho _{cp}^{\ast }=8\rho ab^{2}=\sqrt{2}$.
%Henceforth the asterisks denoting reduced quantities will be omitted. An
%advantage of working with such elongated particles is that the
In this case,
the isotropic-nematic phase transition occurs at quite a low density, 
$\rho /\rho_{\rm cp}\approx 0.2$,
and the simulations are performed at a state point corresponding to
$\rho /\rho_{\rm cp}\approx 0.28$,
for which the nematic order parameter is $Q\approx0.85$.
For this model, temperature is not a significant
thermodynamic quantity, so it is possible to choose $k_{{\rm B}}T=1$
throughout.

The slab geometry is defined by two hard parallel confining walls, which
cannot be penetrated by the {\em centres} of the ellipsoidal molecules.
Packing considerations generate homeotropic ordering at the surface. 
Surface anchoring in this system has been studied recently 
\cite{1999.b} 
for a system with wall separation $L_{z}=125b=8.33a$,
by applying an orienting perturbation at one of the walls
and observing the response at the other.
This yielded an estimate of the extrapolation length
$\lambda\approx 35b \approx 2.33a$.
In the current work,
simulations were carried out for systems 
of $N=2000$ particles with wall separations 
$L_{z}=6.58a,8.22a,9.86a$,
which (from the above estimate of $\lambda$)
would correspond to surface anchoring parameters
in the range $2.8\leq\xi\leq4.2$.

To be sure that we have the same state point for each simulation,
we adjusted the density to have the same $P_{zz}$ component 
of the pressure tensor for all systems. 
Then a sequence of runs was carried out using the constant-$NVT$ ensemble, 
allowing typically $10^{5}$ MC sweeps for equilibration 
and $10^{7}$ sweeps for accumulation of averages
(one sweep is one attempted move per particle).
\section{Simulation results and discussion}
\label{sec:results}
\begin{figure}
\includegraphics[width=8cm]{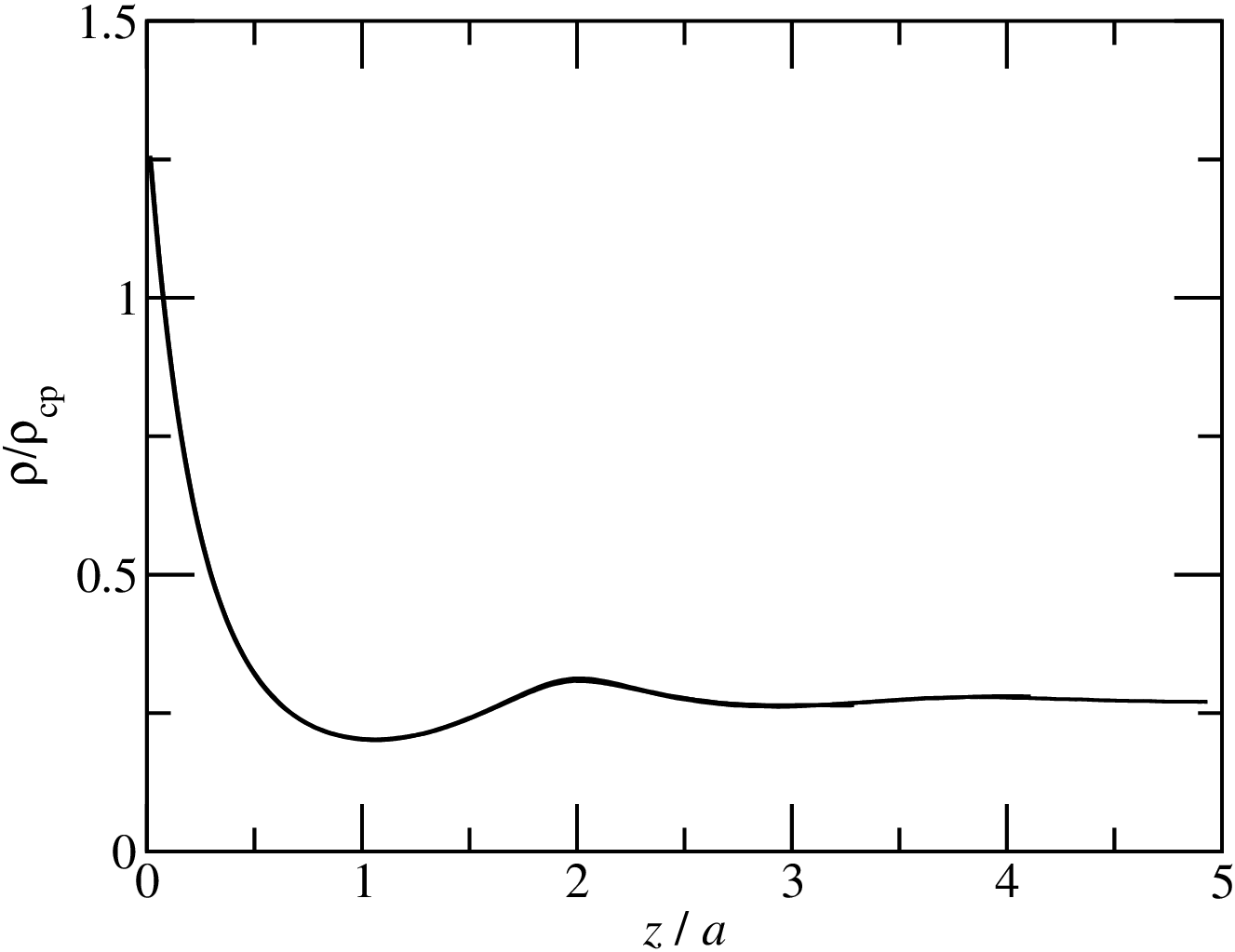}
\caption[Fig 1]{
\label{fig:1} % fig1.ps
Density profiles as functions of $z$-coordinate. 
Distances are normalized by the molecular semi-axis length $a$
and the density is expressed relative to 
the closed-packed density $\rho_{cp}$ for perfectly aligned ellipsoids. 
The profiles are symmetrical, only one side of the slab is shown.
The results for different wall separations are almost indistinguishable
on this scale.
}
\end{figure}

\begin{figure}
\includegraphics[width=8cm]{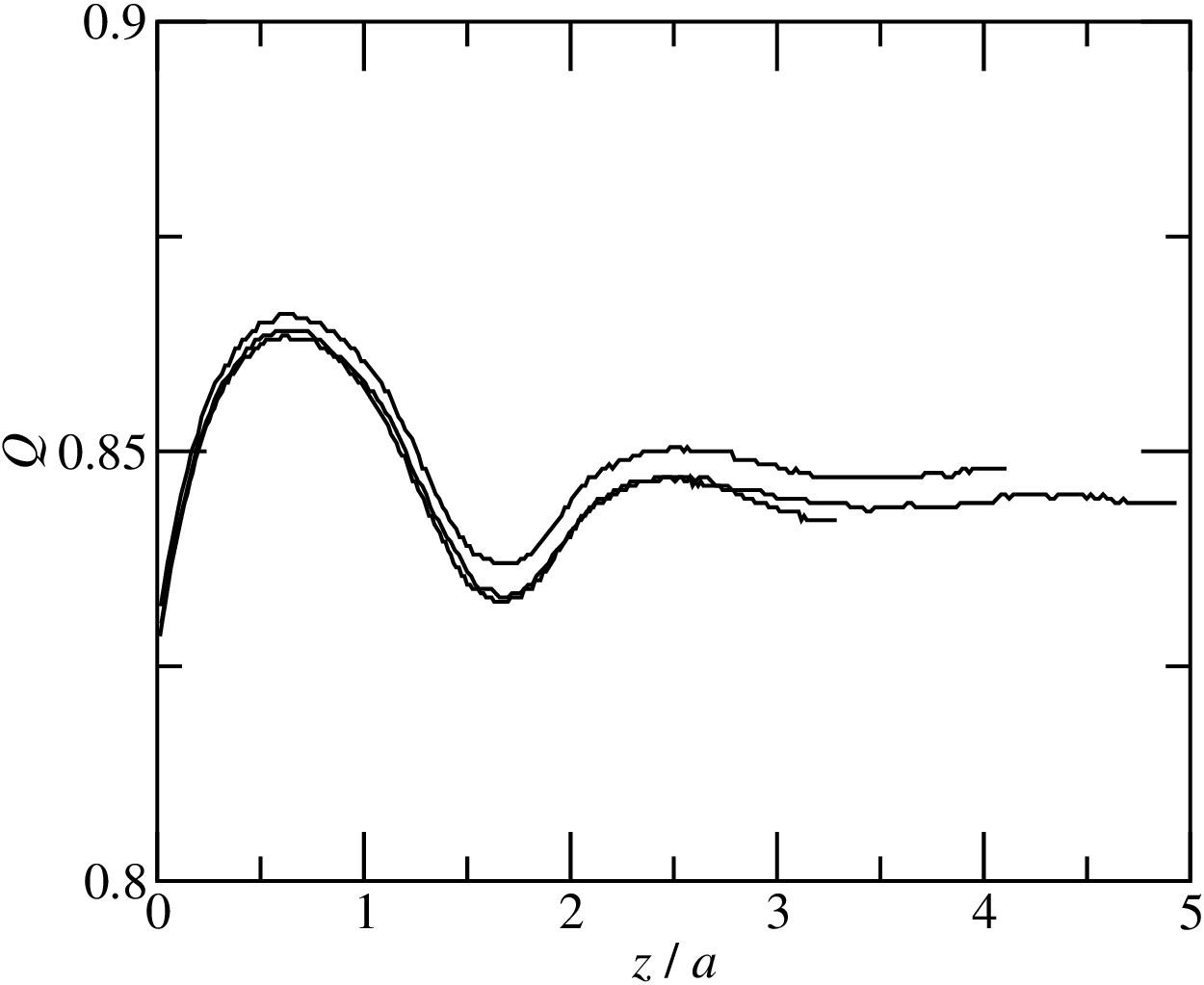}
\caption[Fig 2]{
\label{fig:2} % fig2.ps
Profiles of nematic order parameter $Q$ for different wall separations. 
Distances are normalized by the molecular semi-axis length $a$.
The profiles are symmetrical, only one side of the slab is shown.
Note the highly expanded vertical scale.
}
\end{figure}

The simulation results were analysed to give a density profile, 
and an order tensor profile, 
which are shown in Figs~\ref{fig:1}, \ref{fig:2}. 
From these profiles we can see
that the walls are sufficiently well separated, 
and the variation of the order parameter across the slab is small, 
even in spite of the large change in local density near the walls.

The order tensor fluctuations in reciprocal space were calculated using
expression (\ref{eqn:simul}). To fit the simulation results with the 
elastic theory we have to remember that the size of the simulation 
box $L_{z}$ is not necessarily equal to
the liquid crystal cell thickness $L$ appearing in the elastic theory.
The former is a physical quantity, which, in statistical mechanics,
is determined by the positions at which the liquid number density 
becomes identically zero.
The latter appears in the elastic theory:
it is determined by the positions at which the orientational elastic boundary
conditions are applied.
Physically, the difference may be ascribed to
partial penetration of the walls by the liquid crystal molecules,
formation of an ordered (or solid) layer near the walls,
or other molecular-scale features.
We assume that we may write 
$L=L_{z}+2L_{{\rm w}}$,
where the value $L_{{\rm w}}$ (which may be positive or negative)
is characteristic of the wall,
independent of $L_z$,
and may be determined in our fitting process.

\begin{figure}
\includegraphics[width=8cm]{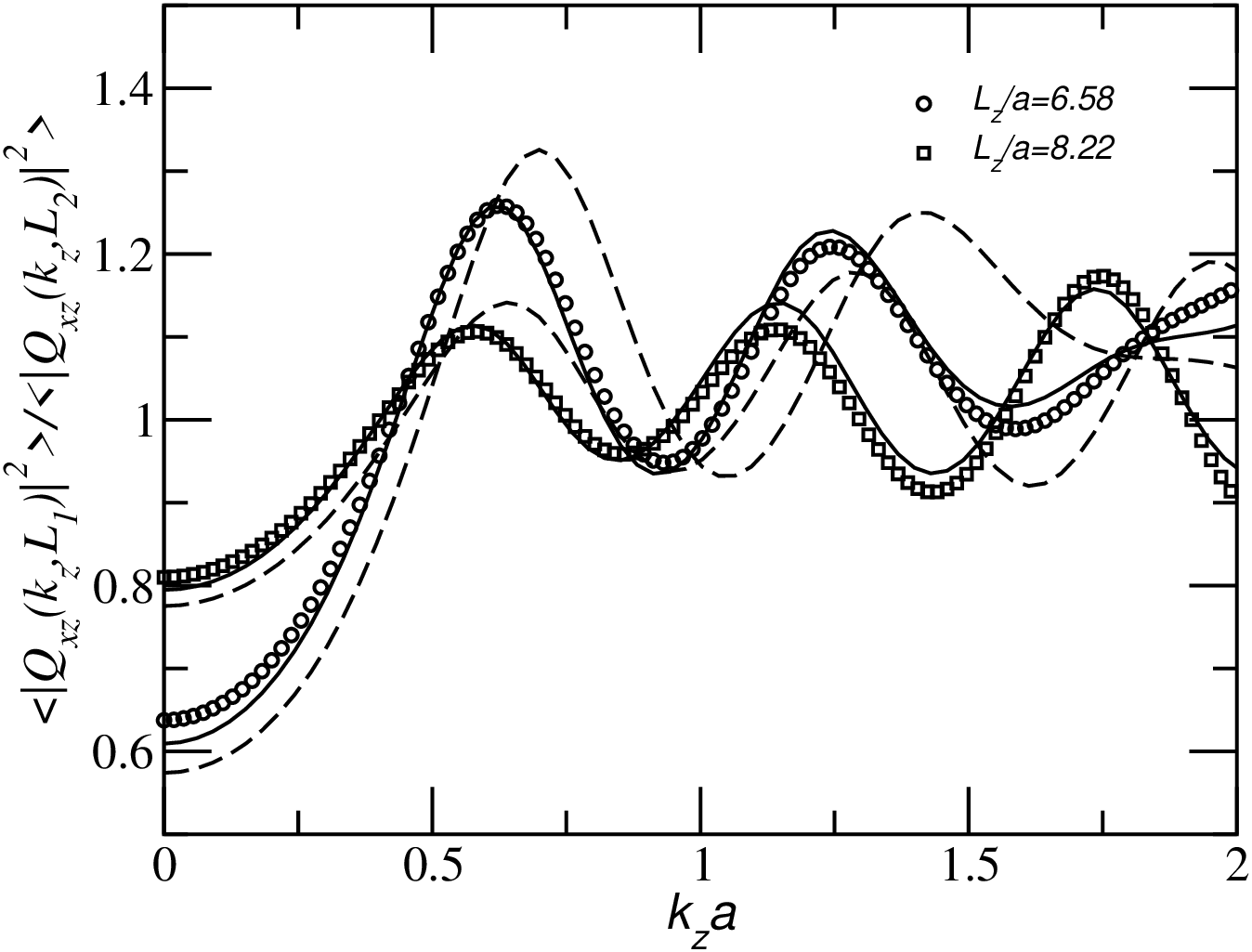}
\caption[Fig 3]{
\label{fig:3} % fig3.ps
Ratio of the director fluctuations as function of wavevector
(normalized by the molecular semi-axis length $a$)
for different wall separations.
Symbols: Monte Carlo results;
solid lines: elastic theory;
dashed lines: elastic theory, without correction for the difference
between the elastic-theory cell thickness $L$  
and the simulation box size $L_z$.
}
\end{figure}

The best estimate of the wall-induced separation distance $L_{{\rm w}}$ 
was obtained examining the ratios $\left\langle \left| Q_{\alpha
z}(k_{z},L_{1})\right| ^{2}\right\rangle /\left\langle \left| Q_{\alpha
z}(k_{z},L_{2})\right| ^{2}\right\rangle$ for different $L_{1,2}$ since 
they reveal more structure
for large $k_{z}$.
Plotting the results in this way
removes the overall scaling of the amplitudes of fluctuations,
which are sensitive to changes in $\lambda$,
while the shapes of the curves, and the characteristic
oscillation wavelengths, are sensitive to the choice of $L_{{\rm w}}$.
These ratios with $L_1 = 6.58a, 8.22a$, 
$L_2 = 9.86a$ and corresponding fitting curves are 
plotted in Fig.~\ref{fig:3}. The best fits were obtained with 
$L_{{\rm w}}/a=0.59$. 
In this figure
we also plot theoretical curves with 
$L_{{\rm w}}=0$:
the clear discrepancy with the simulation results indicates that the
simulation box size $L_{z}$ is indeed significantly different
from the actual cell thickness.

\begin{figure}
\includegraphics[width=8cm]{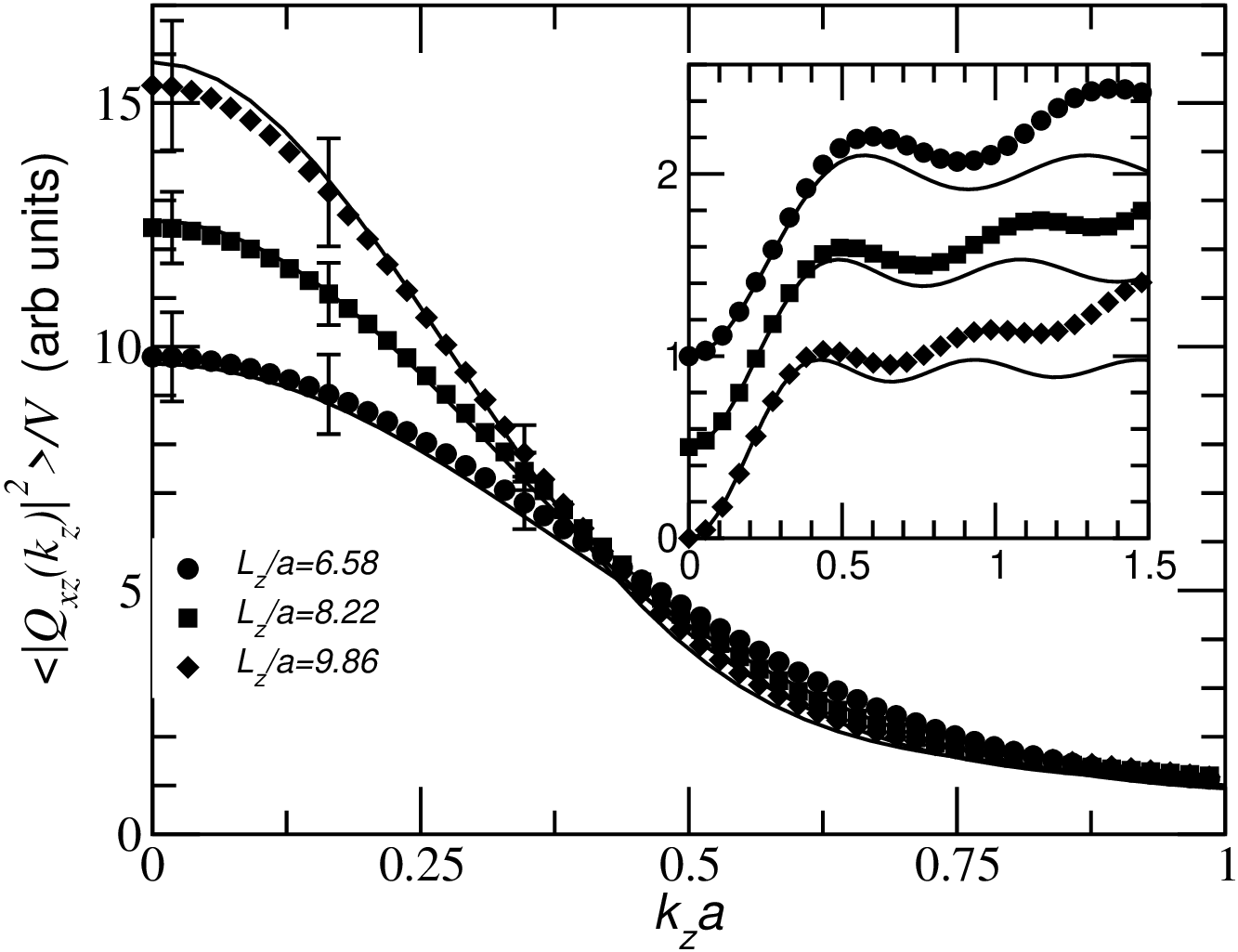}
\caption[Fig 4]{
\label{fig:4} % fig4.ps
Director fluctuations (arbitrary units) as function of wavevector
(normalized by the molecular semi-axis length $a$)
for different wall separations.
Symbols: Monte Carlo results; 
error estimates are indicated at some representative points,
at higher wave-vectors the errors are smaller than the plotting symbols. 
Solid lines: elastic theory, 
fitted to parameters discussed in the text.
Inset: fluctuations multiplied by $(k_za)^2$ to emphasize
structure at higher wavenumbers.
Successive curves are offset by 0.5 for clarity.
}
\end{figure}

Following the determination of $L_{{\rm w}}$,
we adjusted the extrapolation length $\lambda$
to obtain the best fit to the fluctuation data:
the fluctuation amplitudes with
small $k_{z}$ are most sensitive to this quantity. 
Together with the corresponding fitting curves for 
the different slab thicknesses, our results are plotted in Fig.~\ref{fig:4}.
The best fits were obtained with a bulk elastic constant
$Ka/k_{{\rm B}}T\approx66$
and an extrapolation length $\lambda/a\approx2.3$.
The theoretical fitting curves agree well with the simulation results,
for small values of $k_{z}$,
as one would expect for a theory valid for long wavelength fluctuations.
At higher $k_{z}$, the structure (emphasized in the inset
of Fig.~\ref{fig:4} by a multiplying factor $k_z^2$)
is not perfectly reproduced,
but the agreement is satisfactory.
This is not surprising, since
we expect the elastic theory to become less accurate at higher $k_z$.
Finally, we note that the extrapolation distance relative to the
simulation wall position is $L_{{\rm w}}+\lambda\approx2.89a$,
which compares moderately well with the value $\lambda\approx2.33a$ obtained
in the previous study of this system \cite{2000.d}.
It should be noted that the director configuration of that work
does not allow one to determine, separately, $L_{{\rm w}}$ and $\lambda$,
so the quoted value of $\lambda$ really represents $\lambda+L_{{\rm w}}$.

We have to point out some possible limitations of the method. 
The first one is computational time.
The effect of the surfaces is to dampen the amplitude of
long-wavelength modes;
these have the longest relaxation times, 
according to eqn~(\ref{eqn:time}),
and so it is essential to carry out very long runs to adequately
sample them.
We have paid some attention to estimating the error bars on
the measured values of
$\left| Q_{\alpha z}\left( k_{z}\right) \right|^{2}$,
as indicated in Fig~\ref{fig:4}:
the larger values at low $k_z$ follow directly from this effect.
We shall return to examine the time-dependence of fluctuations
in a future publication.
The second limitation is the actual sensitivity of the measured averages 
to the variation in the anchoring strength and cell thickness. 
One might expect that in practice it is
not possible to measure large values of the anchoring parameter 
$\xi=L_z/\lambda$, so we need reasonably thin cells.
As the cell thickness $L_z$ becomes large,
the fluctuation spectrum becomes insensitive to $L_{\rm w}$.
However, it is important that the walls do not become too close:
the {\em bulk} region should be sufficiently large 
compared to the {\em interfacial} region. 
Only in this case can we assume that the scalar order parameter $Q$ 
in the liquid crystal bulk 
is constant for large scale fluctuation modes.

To summarize, analysis of the director fluctuations in 
nematic liquid crystal slabs allowed us to measure the 
surface anchoring strength parameter. 
The method has been tested for a system of 
hard ellipsoids of revolution of elongation $e=15$ 
confined between hard walls with homeotropic anchoring.
Careful analysis of fluctuations in slabs 
of different thickness has allowed us to resolve,
for the first time,
the position of the elastic boundary condition relative
to the simulation wall, as well as measuring the extrapolation length.
The elastic theory gives a good description at low wavenumbers,
but is less accurate at higher wavenumbers.

%\section*{Acknowledgements}
\acknowledgements
This research was supported by EPSRC. 
D.A. acknowledges support through
grant ORS/99007015 of the Overseas Research Students Award; 
G.G. acknowledges the support of a British Council Grant; 
M.P.A. is grateful to the Alexander von Humboldt foundation. 
Some of this work was carried out while visiting
the Max Planck Institute for Polymer Research, Mainz,
and the Institute of Physics, University of Mainz;
the authors are grateful to
Kurt Kremer and Kurt Binder for their hospitality,
and
conversations with F. Schmid and H. Lange are gratefully acknowledged.

\appendix
\section*{Fluctuation spectrum}
\label{sec:appendix}
To obtain the spectrum of the $q_{z}$ modes of the fluctuations 
we need to select from the eigenfunctions 
of the operator $-\frac{1}{2}K\nabla ^{2}$
those which satisfy the boundary conditions (\ref{eqn:boundary}). 
Substituting eqn~(\ref{eqn:fluct}) 
into the boundary conditions (\ref{eqn:boundary}),
we obtain a set of linear equations 
\begin{eqnarray*}
(\xi+{\rm i}\chi) {\rm e}^{{\rm i}\chi}\delta {\bf n}^{(+)}
+
(\xi-{\rm i}\chi) {\rm e}^{-{\rm i}\chi}\delta {\bf n}^{(-)} &=& 0 \\
(\xi-{\rm i}\chi) \delta {\bf n}^{(+)}
+
(\xi+{\rm i}\chi) \delta {\bf n}^{(-)} &=& 0 \:.
\end{eqnarray*}
This set of linear homogeneous equations for $\delta {\bf n}^{(\pm)}$ 
has a nontrivial solution if its determinant equals zero. 
This condition leads to the secular equation 
for the $q_{z}$ vector (\ref{eqn:secular}) 
and relation (\ref{eqn:nplusminus}).

In the case of strong anchoring, $\xi\rightarrow\infty $, 
the $q_{z}$ spectrum is equidistant: 
$q_{z}L=\pi n$, where $n$ is a positive integer. 
For finite anchoring coefficient $\xi $, we have a shift in this
spectrum.
The magnitude of the shift depends on the anchoring parameter $\xi $. 
Indeed, for sufficiently strong anchoring parameters, $\xi >>1$,
asymptotically: 
\[
q_{z}L=\pi n-\frac{2\pi n}{\xi },\quad n=1,2,...,\quad n/\xi <<1,
\]
which is equivalent to replacing $L$ by $\left( L+2\lambda \right) $, 
$\lambda $ being the extrapolation length.

For weak anchoring, $\xi <<1$: 
\begin{eqnarray*}
q_{z}L &=&\pi n+\frac{2\xi }{\pi n},\quad n=1,2,..., \\
q_{z}L &=&\xi ^{1/2},\quad n=0 \:. 
\end{eqnarray*}

The spectrum of the $q_{x}$,$q_{y}$ wavevectors depends on the system
geometry. 
Again, 
if we have periodic boundary conditions in $x$ and $y$ directions, 
the $q_{x}$ and $q_{y}$ have a discrete spectrum, on a fine grid 
$q_{\alpha }L_{\alpha}=2\pi n_{\alpha }$,
$(n_{\alpha }=0,1,2,...,)$, otherwise 
${\bf q}_{\bot }=(q_{x},q_{y})$ are unrestricted.

It is also easy to show that the eigenfunctions 
which correspond to different eigenvalues $q_{z}$ and $q_{z}^{\prime }$ 
are orthogonal. 
Indeed,
using eqn.(\ref{eqn:nplusminus}) we can rewrite 
\begin{eqnarray*}
\Phi({\bf q}_{\bot },q_{z})  &=&
\delta {\bf n}^{(+)}({\bf q}_{\bot},q_{z}) {\rm e}^{{\rm i}q_{z}r_{z}}
+
\delta {\bf n}^{(-)}({\bf q}_{\bot},q_{z}) {\rm e}^{-{\rm i}q_{z}r_{z}}
\\
&=&
\frac{2{\rm i}}{{\rm i}\chi +\xi }
\left[ \chi\cos(q_{z}z) +\xi\sin(q_{z}z) \right] 
\delta {\bf n}^{(+)}({\bf q}_{\bot},q_{z}) \:.
\end{eqnarray*}
It is easy to check using the secular equation (\ref{eqn:secular}) 
that functions 
$\phi(q_{z})=\chi\cos(q_{z}z) +\xi\sin(q_{z}z)$ 
are orthogonal, i.e.\
\begin{equation}
\int_{0}^{L}\phi(q_{z}) \phi(q_{z}^{\prime }) dz
=
\frac{1}{2}L\left(2\xi+\chi^2+\xi^2\right)\delta_{q_{z},q_{z}^{\prime}}
\:,  
\label{eqn:norm}
\end{equation}
where $\delta _{q_{z},q_{z}^{\prime }}$ is the Kronecker delta. 
Therefore, the eigenfunctions 
$\Phi({\bf q}_{\bot },q_{z}) $ 
are orthogonal and can be normalized using eqn~(\ref{eqn:norm}).

% \bibliography{journals,main,mike} % paper.bbl now included

\newpage

\end{document}